\documentclass[12pt,a4paper]{article}
\usepackage{mathptmx} 
\usepackage{newtxtext,newtxmath} 

\usepackage[left=2.5cm, right=2.5cm, top=2.5cm]{geometry} 
\usepackage{setspace}
\usepackage{endnotes}

\usepackage{graphicx}
\usepackage{multirow}
\usepackage{hhline}
\usepackage{subcaption} 

\usepackage{array}
\usepackage{booktabs}
\usepackage{multirow}
\usepackage{siunitx}
\usepackage{appendix}
\usepackage{color}
\usepackage{amssymb}
\usepackage{pdflscape}
\usepackage{amsmath}
\usepackage{natbib} %

\usepackage{lipsum}
\def\CP{t_{\rm \small CP}}

\makeatletter
\renewcommand{\maketitle}{\bgroup\setlength{\parindent}{0pt}
\begin{flushleft}
  \textbf{\@title}\vspace*{.5cm}

  \@author
\end{flushleft}\egroup
}
\makeatother

\date{}
\author{%
Zane Hassoun$^1$, Ben Powell$^2$, Niall MacKay$^3$ \\
\vspace*{.25cm}
Department of Mathematics, University of York, York, UK }

\title{\huge{\textbf{Kairosis: A method for dynamical probability forecast aggregation informed by Bayesian change point detection}}}

\begin{document}
\onehalfspacing 

\maketitle
\footnotetext[1]{Email: zane.hassoun@york.ac.uk}
\footnotetext[2]{Email: ben.powell@york.ac.uk}
\footnotetext[3]{Email: niall.mackay@york.ac.uk}



\begin{abstract}
We present a new method, ``kairosis", for aggregating probability forecasts made over a time period of a single outcome determined at the end of that period. Informed by work on Bayesian change-point detection, we begin by constructing for each time during the period a posterior probability that the forecasts before and after this time are distributed differently. The resulting posterior probability mass function is integrated to give a cumulative mass function, which is used to create a weighted median forecast. The effect is to construct an aggregate in which the most heavily weighted forecasts are those which have been made since the probable most recent change in the forecasts' distribution. Kairosis outperforms standard methods, and is especially suitable for geopolitical forecasting tournaments because it is observed to be robust across disparate questions and forecaster distributions.
\end{abstract}

\section{Introduction}
\label{sec1}
\begin{figure}[h]
    \centering
    \includegraphics[width=\linewidth]{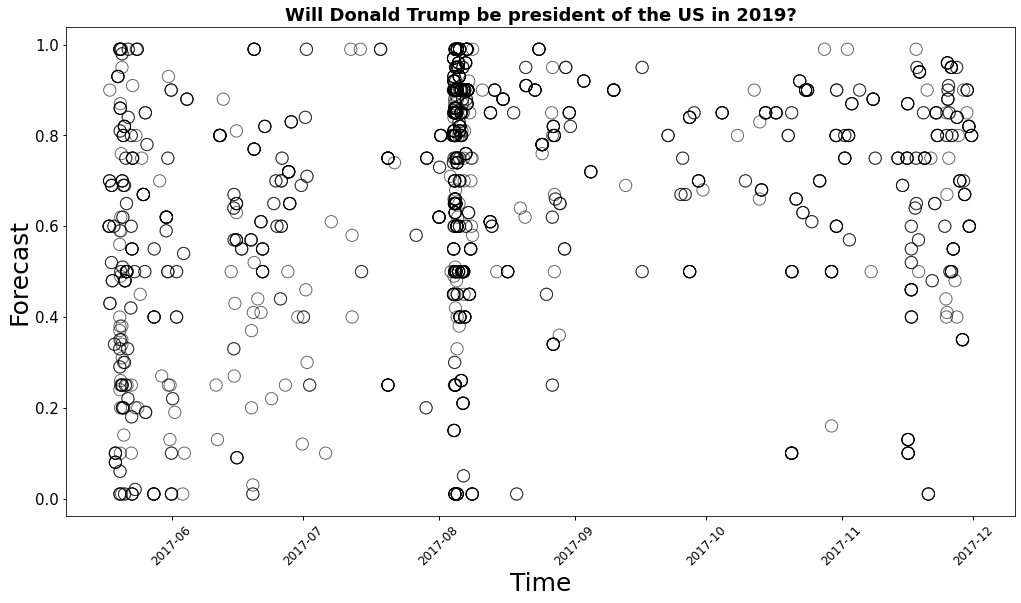}
    \caption{\textbf{Metaculus forecasters predict Trump presidency continuing to 2019}. Probabilistic forecasts, unattached to specific forecasters, are recorded over 7 months in 2017. Forecast values of zero and one correspond to certainty that Trump will and will not be President in 2019, respectively. Such data are the starting point of the kairosis calculations.}

    \label{fig:Trump Example}
\end{figure}

\noindent Geopolitical forecasting tournaments have become increasingly popular over the last decade, notable providers include the Good Judgment Project and Metaculus. A typical question from Metaculus is that of Figure \ref{fig:Trump Example}, ``Will Donald Trump be president of the USA in 2019?".  From when the question opened (May 17, 2017), forecasters submitted probability forecasts (on a scale of $0$ to $1$), until the question was resolved on Feb 1, 2019, although here we show only the first seven months' forecasts. A forecaster may make as many or as few forecasts as they wish, at any time they wish; thus we observe only the overall evolving distribution of forecasts, rather than multiple time-series of individual forecasters' evolving beliefs. 

After resolution, the forecasts are scored. If forecasts are considered ``static", taking no account of {\em when} the forecast is submitted, a simple proper probability score, such as the Brier (quadratic, \citealt{brier1950score}) or Log (logarithmic, \citealt{good1952rational}) score, can be used. Proper scores are optimized by, and therefore incentivize forecasters to submit their best estimates of, the true probability -- although propriety fails if rewards are not proportional to the score, for example if the prize goes to the overall winner (see for example \citealt{roulston2007performance}, \citealt{brocker2007proper}). But prescience is clearly valuable \citep{schuler2025foresight}, and Metaculus, for example, weights the score by how long it was submitted before resolution. Indeed it is clear just from a visual inspection that the distribution of forecasts is not stationary; it changes both smoothly and sharply at certain points, as does the density of forecasts submitted. The reason is obvious: news occurs much in this way, and new information is continually informing the forecasting process.

Forecast aggregation, by which we can access the ``wisdom of crowds" \citep{surowiecki2005wisdom}, is well developed for static probability forecasts. With no information beyond the raw distribution of forecasts one can use simple measures of central tendency such as the mean or median, more subtle measures such as the extremized mean \citep{atanasov2017distilling}, or more exotic statistics of the distribution \citep{powell2022skem}. If it is possible to measure information heterogeneity or forecaster quality, much more can be done, typically by constructing various forms of weighted pool \citep{ranjan2010combining,clements2011combining,satopaa2014combining,budescu2015identifying}. Even more is possible when one asks the forecasters about others' likely views \citep{palley2019extracting,prelec2004serum}. For a recent review see \cite{winkler2019perspective}.

In contrast, the study of dynamic forecasting problems such as those of Figure \ref{fig:Trump Example} is in its infancy. Suppose that, at any given (``present") time within the question window, we wish to construct the best possible current forecast from all forecasts already submitted. We know nothing about the knowledgeability of the individual forecasters, or the evidence informing their forecasts. Clearly,  static techniques, which aggregate all extant forecasts regardless of submission time, are inappropriate and would be expected to be sub-optimal. But what should be done instead? The state of the art is summarized by \cite{himmelstein2023wisdom}, who begin with the central assumption that forecasts should tend to improve over time; two suggestions are to discount the past by weighting exponentially, or to select the most recent 20\% of forecasts. Yet these neglect the information which a visual inspection of the plot immediately shows is present, concerning past trends and events: the distribution of forecasts is clearly evolving, with moments of change and intense forecasting. At its simplest, we could instead try to identify the most recent ``change point" (at which the underlying statistical distribution of forecasts changed), and use only forecasts made since then. This effectively assumes that at the change point significant new information became available but thereafter no more emerged.

However, as \cite{himmelstein2023wisdom} note, ``it is possible to envision a hybrid method ... of the selection and weighting approaches", and it is our purpose in this article to provide this.  Our technique is informed by, but not equivalent to, Bayesian change point detection, and can be viewed as something like an interpolation between  exponential discounting and the ``most recent change point" method suggested above. It achieves this by using exponential discounting as its prior (starting) position, and then constructing for each possible change point a likelihood that this splits the forecasts into two sets, before and after the supposed change point, with distinct distributions. This gives us a posterior likelihood for every possible change point, whose probability density is then integrated to give a cumulative mass function, which is used to weight past forecasts. 
The practical effect of our method is to create a weighting with multiple downward (as we move into the past) steps, each corresponding to significant changes in the distribution of forecasts.  At one extreme, if there is no obvious time at which the distribution of forecasts changes, the weighting stays close to the original exponential discounting of past forecasts. At the other extreme a single, obvious change point between very different distributions of forecasts is effectively a horizon, behind which old forecasts add nothing.

It is clear from Figure \ref{fig:Trump Example} that forecasts are not made at a uniform rate in chronological time; rather there are intense bursts of activity, typically due to the publication of relevant news, interspersed by quieter periods. If $R$ forecasts are made at chronological times $t_r$, $r=1,\ldots,R$, then the forecast order $1,\ldots,R$ effectively creates a ``forecaster clock" that dilates and magnifies the moments at which forecasts are being made at a high rate, and it is in this ``forecaster time" that we search for change points.
However, intense forecasting activity need not imply that the information landscape and distribution of forecasters' beliefs are changing. To conceptualize the moments of change, we note that the ancient Greeks distinguished $\chi \rho \acute{o} \nu o\zeta$ ({\em chronos}), chronological or calendar time, from $\kappa \alpha \iota \rho \grave{o} \zeta$ ({\em kairos}), the ``time" of critical moments of lived experience \citep{smith1969time}, and our weighting function can then be thought of as the transformation from chronos to kairos (an example of which is presented in Figure \ref{fig:CDF_Example_Intuit}b).  ``Kairos" has a nice interpretation as ``a moment of time when a prophecy was pronounced" \citep{tzam2007kairos}, which fits comfortably with our conjecture that change points in forecast distributions are being driven by the arrival of new information. ``Kairosis" then makes for a correspondingly concise name for our method.

As noted above, dynamical probability forecast aggregation is a nascent topic whose recent developments are nicely summarized in \cite{himmelstein2023wisdom}. Many of the most prominent involve tracking individual forecasters tracked over time. \cite{himmelstein2021forecasting}, for example, use structured regression models which use smooth exponential or logarithmic functions for time dependence in order to identify skilled forecasters. \cite{regnier2018probability} posits a number of properties a good time-series probability forecast should have and uses them to diagnose improvability and inefficiency. \cite{augenblick2021belief} track individuals and analyse the rationality of their Bayesian updates. \cite{satopaa2014agg} use a hierarchical model to analyse infrequently updated forecasts, intractable by standard time-series techniques. In a slightly different context \cite{wawro2014designing,wawro2022time} advocate structured regression models and Bayesian change point detection for historical analysis of time series.

In contrast, in the Metaculus tournament data we consider in our current work, individual forecasters are not identifiable and may make as many or as few forecasts as they wish. Our problem is essentially to extract the best possible current estimate of the crowd view from the dynamics of the evolving distribution of forecasts. Although it is not currently the focus of our attention, it would be perfectly possible to develop our method to incorporate additional weights based on forecasters' skill. Similarly, kairosis could be modified to work with other aggregation techniques from the static wisdom-of-crowds literature -- using, for example, different measures of central tendency of the kairosis-weighted probability distribution, such as an extremized mean \citep{baron2014two} or its skew-adjusted variant \citep{powell2022skem}.

The article is structured as follows. In Section \ref{Algo} we describe our method in detail, illustrating in Figures \ref{fig:Trump Example}--\ref{fig:CDF_Example_Intuit} its application to the question of the Trump presidency in 2019. In Section \ref{sect:results}, after establishing a set of metrics and comparators, we report empirical results across 650 forecasting problems posed by Metaculus. The main part of the paper finishes with a discussion of possible developments and applications. In \ref{simmed_example_sect} we provide an additional illustrative example of computing a forecast-weighting function with our kairosis methodology in order to cement the ideas presented in the body of the paper. In \ref{non_prob_appendix} we present the preliminary results of the extension of kairosis to point forecasts of continuous variables. In \ref{sensitivity_appendix} we report the results of a sensitivity analysis that accompanies the findings of Section \ref{sect:results}.

\section{Methods}\label{Algo}

In this section we expand on our proposed method for aggregating subjective probability forecasts and explain the calculations involved. The result is a set of aggregation weights derived from the CMF of a posterior distribution over times at which significant changes in forecaster behaviour are thought to occur.

\subsection{Deriving a distribution over change point locations}\label{deriving_distr_sect}

Formally, we use Bayes' theorem to obtain a posterior mass function for the time of the most recent change point. In this model there is a kairos, a most recent critical event or change point which actually occurred at $\CP$, and we wish to construct the likelihood (a probability mass density which is a function of time) that this critical event occurred at any given candidate time $t$. The corresponding CMF provides us with the posterior probability that the most recent change point occurred earlier than the given $t$. Equivalently, this is the posterior probability that a forecast immediately following $t$ was made after the most recent change point and so ought to contribute to our post-change point aggregated forecast, while those made before $t$ should not.

To make the case more formally we can introduce a binary weighting function
\begin{align}
w(s \mid \CP=t) = 
\begin{cases}
0 & s<t,\\
1 & s \geq t
\end{cases}
\end{align}
that tells us how to weight a forecast made at time $s$ conditional on the most recent change point's having occurred at time $t$. Then, taking an expectation over possible values of $\CP$, we are led to the expression
\begin{align}
\mathbb{E}_t\left[ w(s \mid \CP=t) \right] := \int_{-\infty}^{\infty} w(s \mid \CP=t) P(\CP=t) \ dt = \int_{-\infty}^{s} P(\CP=t) \ dt
\end{align}
which is exactly the CMF for $\CP$ evaluated at $s$.

Now, returning to our central example, suppose it is October 10, 2017. Our data are $R$ forecasts made at times $t_r$, $r=1,\ldots,R$, up to this date and we are tasked with submitting an optimal aggregated probability forecast to answer the question of the 2019 US President. It is reasonable to assume that most of the data is relevant, but to what extent is not immediately apparent. One might, for example, propose that more recent data is likely to be more informative than that from the distant past. On the assumption that critical events invalidating preceding forecasts occur independently at a constant rate over time, we are led naturally to the idea of exponentially discounting old forecasts.      

We therefore begin with a prior assumption of a constant probability $p \in [0,1]$ that at least one change point-inducing event occurs between consecutive forecasts. This motivates a geometric prior distribution on the time of the last change point, $P(\CP=t_r) = p (1-p)^{R-r}$, which follows from the idea that for $t_r$ to have been the last change point there must have been $R-r$ subsequent inter-forecast periods without a change point. We then update our prior distribution given knowledge of observed forecasts using Bayes' theorem,
\begin{align}
P(\CP=t_r|\text{Forecasts}) = \frac{{P(\text{Forecasts}|\CP=t_r)}{P(\CP=t_r)}}{P(\text{Forecasts})}.\label{bayestheorem1}
\end{align}

The next key ingredient for our methodology is thus to specify the distribution of the forecasts given a candidate change point. Inspired by work on Bayesian hypothesis testing by \cite{holmes2015two}, we use the compound Dirichlet-categorical distribution to describe the number of forecasts falling in different sub-intervals (``bins") of $[0,1]$. We use the probabilities this distribution assigns to the observed bin-counts to inform the quantity $P(\text{Forecasts}|\CP=t_r)$ appearing in \eqref{bayestheorem1}.

\begin{figure}[h]
    \centering
    \includegraphics[width=\linewidth]{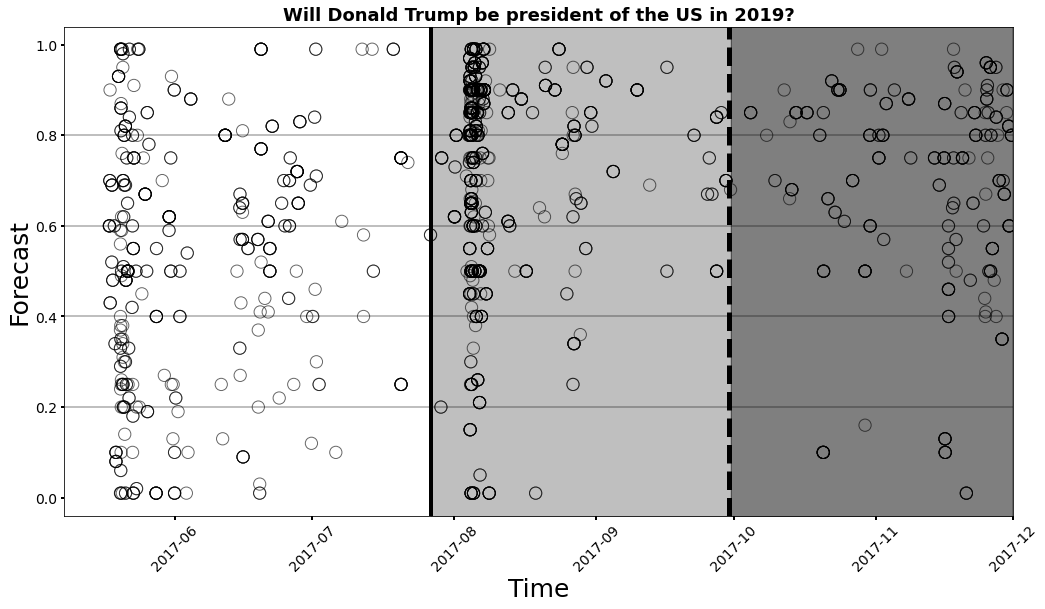}
    \caption{\textbf{Forecasts are partitioned to measure their relative frequencies during different intervals}. A forecast aggregator needs to make her own forecast at Oct 10, 2017 (dashed vertical line). She is unaware of forecasts in the darkest shaded region. She considers the relative probabilities of change points having occurred immediately before each of the individual forecasts were made. The solid vertical line illustrates one candidate. Her calculations are based on relative frequencies of forecasts in a set of bins (separated by horizontal lines).}
    \label{fig:dirichlet-discrimination}
\end{figure}

The Dirichlet-categorical distribution is a compound distribution from which we can produce samples in two stages: first, drawing latent probabilities from an underlying Dirichlet distribution, then using these probabilities to assign labels to objects. The conjugacy between the Dirichlet and categorical distributions make it a natural choice for Bayesian modelling of categorical data and, importantly, leads us to a closed-form expression for its probability mass function.

The Dirichlet-categorical distribution assigns probability mass
\begin{equation}\label{DCmassfun}
P(n_1,\ldots,n_K)=\frac{\Gamma(\sum \alpha_k)}{\Gamma(\sum n_k+\alpha_k)}\prod_{k=1}^{K}\frac{\Gamma(n_k+\alpha_k)}{\Gamma(\alpha_k)},
\end{equation}
to the event in which bin counts $n_k$ are observed for bin labels $k=1,\ldots,K$.
The gamma function, an analytic extension of the factorial, is  $\Gamma(x)=\int_0^\infty s^{x-1}e^{-s}\,ds$. The $\alpha_k \geq 0$ are parameters for the Dirichlet distribution that is effectively averaged over. They are usefully interpreted as pseudo-counts, reflecting approximate \textit{a priori} beliefs for the true values of the bin probabilities -- thus, as we shall require later, small values of $\alpha_k$ correspond to uninformed priors, while large $\alpha_k$ impose strong, informative prior beliefs. For readers familiar with information entropy, it is interesting to note that in the limit in which all the counts become very large we can, using Stirling's formula, derive 
\begin{equation}\label{DCmassfun2}
\lim_{n_1,..,n_K \rightarrow \infty} \log P(n_1,\ldots,n_K) = N \sum_{k=1}^{K} \frac{n_k}{N} \log \left( \frac{n_k}{N} \right)
\end{equation}
where $N=\sum_{k=1}^{K}n_k$. We recognize this as being proportional to the negative entropy for the sample distribution of forecasts across bins, equivalently its Kullback-Leibler divergence from the uniform distribution. The implication here is that, certainly for large $n_k$, the Dirichlet-categorical distribution assigns most mass to outcomes with low entropy, in which most forecasts fall in a small number of bins. Informally, we might say that the distribution anticipates agreement among forecasters to the extent that their forecasts concentrate on a small subset of possible values.

Now, to propose a change point at time $\CP=t_r$ (that is, just after the $r$th forecast) is to suppose that forecasts up to and after $t_r$ follow two independent Dirichlet-categorical distributions, because an event is thought to have occurred just after $t_r$ that has fundamentally changed the forecasters' (unobserved) distribution of beliefs. Of course this is never wholly true -- forecasts always incorporate the previously-available information which informed forecasts before $t_r$ -- but we now have a way of quantifying the probability that it is so. Thus, for a single change point at $t_r$, expression \eqref{bayestheorem1} becomes
\begin{align}
\hspace*{-0.2in}P(\CP=t_r|\text{Forecasts}) 
\propto& {P(\text{Forecasts}|\CP=t_r)}{P(\CP=t_r)} \notag \\
=&\frac{\Gamma(\sum \alpha_k)}{\Gamma(\sum n_k+\alpha_k)}\prod_{k=1}^{K}\frac{\Gamma(n_k+\alpha_k)}{\Gamma(\alpha_k)}  \notag \\
& \times \frac{\Gamma(\sum \alpha_k')}{\Gamma(\sum n_k'+\alpha_k')}\prod_{k=1}^{K}\frac{\Gamma(n_k'+\alpha_k')}{\Gamma(\alpha_k')} \notag \\
& \times p(1-p)^{R-r} \label{bayestheorem2}
\end{align}
where the unprimed and primed letters denote counts $n_k$ and pseudo-counts $\alpha_k$ up to and after $t_r$.

We then evaluate \eqref{bayestheorem2} for every candidate change point $r=1,\ldots,R$, taking us from calendar time 17 May 2017 to 10 Oct 2017 in our example. With each evaluation, we are asking the question ``What is the probability that this set of forecasts is actually drawn from two different distributions, one up to and another after our candidate $t_r$?" A visualisation of a single step in the process is shown in Figure \ref{fig:dirichlet-discrimination} where the two periods are highlighted with no shading and light gray shading, respectively. Having computed \eqref{bayestheorem2} for each time point, we can normalize it to derive our posterior mass function for the location of the change point. The corresponding CMF then provides weights for our aggregated forecast.
For readers who might find it helpful, \ref{simmed_example_sect} provides an illustrative example of how kairosis works on simulated data with a simplified set of just three possible change points.

\begin{figure}[htbp]
    \centering
    \begin{subfigure}{1\textwidth}
        \includegraphics[width=\linewidth]{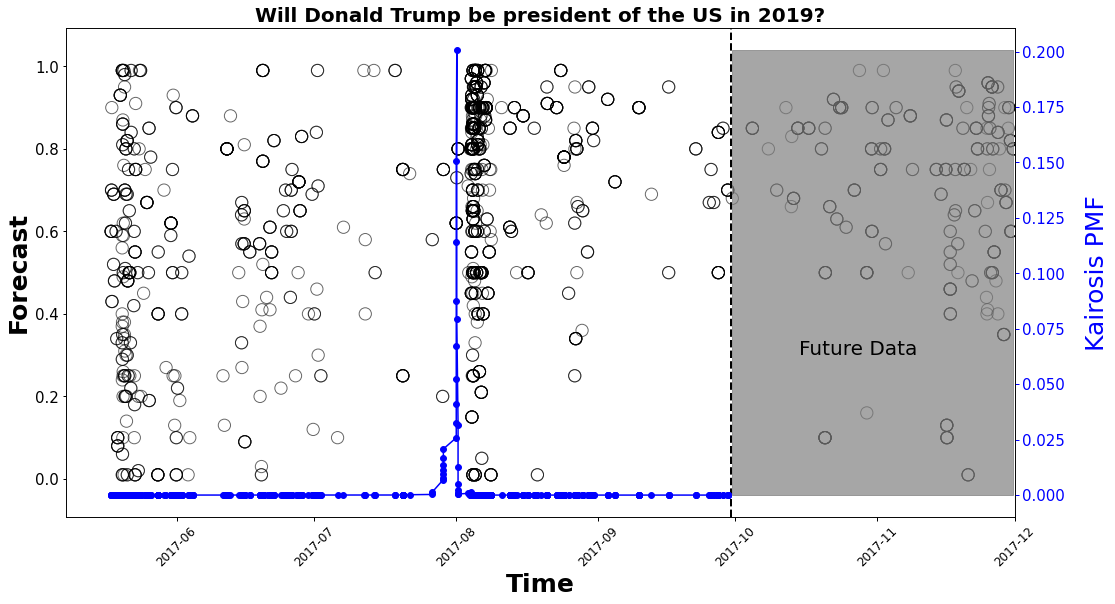}
        \caption{}
        \label{fig:sub1}
    \end{subfigure}
    
    \begin{subfigure}{1\textwidth}
        \includegraphics[width=\linewidth]{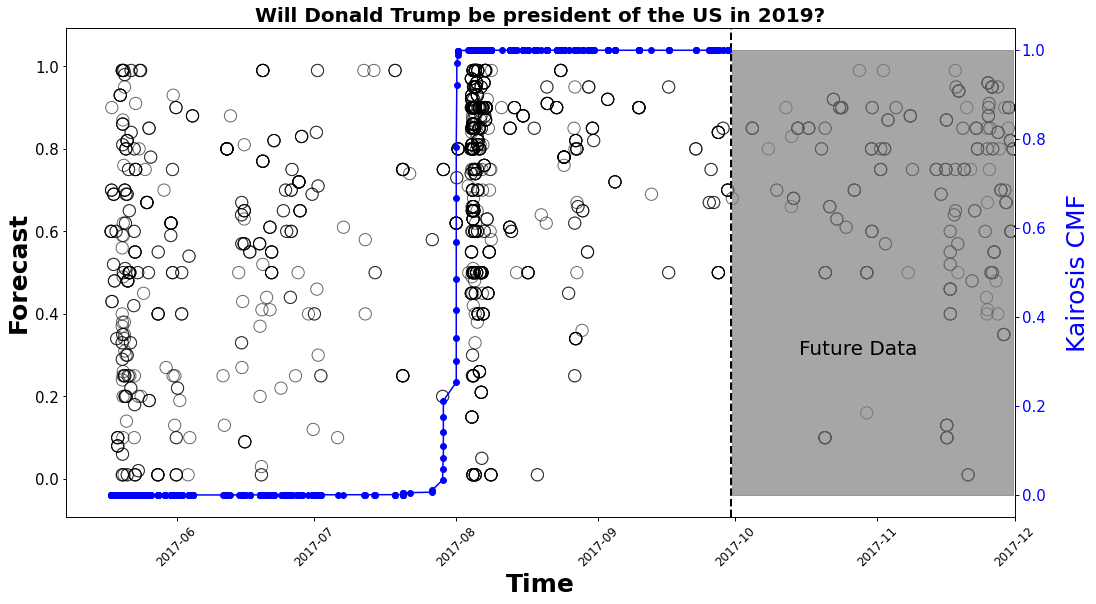}
        \caption{}
        \label{fig:sub2}
    \end{subfigure}
   
    \caption{\textbf{Kairosis weights are computed from the cumulative mass function (CMF) of the posterior distribution for the most recent change point}. Values are calculated for (a, upper) the mass distribution function and (b, lower) the cumulative mass function for the posterior distribution over potential times for the most recent change point. They are interpolated and superposed on Subfigures \ref{fig:sub1} and \ref{fig:sub2}, respectively, with their vertical axis labels on the right of the plots. For the CMF this happens to be the same $[0,1]$ interval as for the forecasts, but the PMF is on a dissociated scale from $0$ to just above its maximum value. The CMF then provides the kairosis weights for aggregating the forecasts.}

    \label{fig:CDF_Example_Intuit}
\end{figure}

The mass and cumulative mass functions for our running example are illustrated in Figure \ref{fig:CDF_Example_Intuit}. The dominant feature here is a flurry of forecasting activity in early August 2017. The ``clock" of forecaster time is running fast here, increasing the concentration in chronological time of potential change points. Our posterior probability for a change point accounts for this activity and, using the Dirichlet-categorical term in \eqref{bayestheorem2}, considers whether the distribution of forecasts actually changes here. Indeed, it appears to shift upwards, and the probability of a change point is deemed to be high. The resulting CMF features a large, sudden rise with the effect that our aggregation weights also rise. With the vertical scale from 0 to 1 as shown, all forecasts after early August receive weights between 0.8 and 1.0, while those before are given weights less than 0.25. This relative downweighting of earlier forecasts by a factor of 3 to 4 is then kairosis's belief of what is appropriate given its degree of confidence in the time of the most recent change point. 

(It is perhaps worth noting that the timing of this probable change point coincides with the announcement of the handing-over, from the Trump campaign team to the US Senate Judiciary Committee, of documents relating to suspected Russian collusion in the 2016 presidential election. Although the link between the announcement and the probable change point obviously cannot be ascertained here, it does provide a plausible explanation for the data.)

In addition to their primary use as weighting functions, we find it useful to interpret CMFs such as those illustrated in Figure \ref{fig:CDF_Example_Intuit}b as functions nonlinearly transforming chronological time to a scaled version of our notional kairos time. This concept guides our intuition when reconciling contextual information, observed forecaster data and aggregation weights.

\subsection{Parameter Selection}

The calculations described in the preceding section rely on the specification of a small number of parameters and modelling choices:
\begin{enumerate}
    \item the binning of $[0,1]$ from which the counts $n_k$ are derived,
    \item the change point occurrence rate $p$, and
    \item the $\{\alpha_k,\alpha'_k\}$ that parameterize the prior distributions for the bin probabilities before and after a putative change point.    
\end{enumerate}

When considering the specification of forecast bins, we emphasize that the binning is only for the purpose of assessing the likelihoods of a change point. Once the change point CMF has been constructed, it is used to weight the  original, precise forecasts. In our numerical experiments we have chosen to partition $[0,1]$ into five equally-sized intervals, representing a fairly coarse-grained discretization of the forecasts.

Remembering that our forecaster clock is ticking with the arrival of each forecast, we can think of the prior change point occurrence rate $p$ in terms of its reciprocal $1/p$, which describes the expected number of forecasts per change point. Seen this way round, the parameter can be understood as describing the responsiveness of the population of forecasters, which in the Metaculus data remains fairly constant from question to question even though the question topics may vary considerably. In the numerical comparisons of Section \ref{sect:results} we have specified $1/p=10$, meaning that we expect a change point to motivate approximately 10 forecasters to contribute updated forecasts. Sensitivity analyses presented in \ref{sensitivity_appendix} show that this choice generally leads to near-optimal Brier scores for the kairosis-aggregated forecasts and for the aggregated forecast with geometrically decaying weights -- beyond the avoidance of very low values, there is no need for fine tuning of $1/p$ to achieve good results. We note also that the prior distribution for the change point location tends towards a uniform distribution as $1/p$ becomes very large so that all values of $1/p \gg R$ lead to approximately the same posterior inferences.

Recall that the $\alpha_k$ and $\alpha_k'$ are pseudo-counts quantifying our prior beliefs for the proportion of forecasts falling in each bin, to either side of a putative change point. They appear in the likelihood function for the change point (equation \eqref{bayestheorem2}) via the quantities $\alpha_k+n_k$ and $\alpha_k'+n_k'$. First, consider the situation {\em after} the most recent change point. The forecasts here are understood to come from a single distribution and we expect this distribution to have low entropy, meaning that forecasts are concentrated in a small number of bins. When this is observed to be the case for a candidate change point, then our posterior probability for its being the most recent change point is increased. Small values of $\alpha_k'$, equal for all bins, correspond to a prior that allocates high probability to such low-entropy forecast distributions, and we choose $\alpha'_k=1$. Next, consider the situation {\em before} the most recent change point. This can be more complicated, since the forecasts here could have resulted from multiple inter-change point distributions. We cannot expect them to come from a single distribution characterized by high levels of agreement among forecasters. We therefore encode the anticipated diversity of historical forecasts, which grows as that history increases in duration, by setting $\alpha_k = \lambda \CP$ where $\lambda>0$ is a positive scaling factor. An effect of this choice is to limit, but not ignore, the influence of distant history on our estimate for the most recent change point. In the calculations of Section \ref{sect:results} we have chosen to set $\lambda=0.2$. This choice is motivated by the idea that for every forecast in the distant past (i.e. before the most recent change point) we add another pseudo-forecast that is effectively spread out over the five bins. Our sensitivity analyses indicate a high level of robustness to choices of $\lambda>0.1$.

The overall results of the sensitivity analysis are conveyed by Figure \ref{fig:sensitivity_analysis}. To summarize: other than the avoidance of very high values of $p$ (corresponding to a prior belief in very frequent change points) and very low values of $\lambda$ (almost no stabilization of pre-change-point forecasts) there is no need for fine tuning of $p$ or $\lambda$ in order for kairosis to achieve its results.

\section{Results}\label{sect:results}

To test our kairosis methodology we study its performance for 650 Metaculus forecasting questions. These questions vary both in the length of time the question is open and in the number of forecasts received, with a mean of 145 days and 893 forecasts. The questions span diverse topics, including international conflict ({\em e.g.\ }``Will Russia expand by means of armed conflict before 2020?"), energy ({\em e.g.\ }``Will radical new  low energy nuclear reaction  technologies prove effective before 2019?"), and business and finance ({\em e.g.\ }``Will there be a financial crisis in China in 2017?"). 

The mean and median forecasts across the 650 questions were $0.436$ and $0.444$, respectively. The standard deviation for the forecasts (computed for each question and then averaged over questions) was $0.181$, quantifying the level of agreement typically seen between forecasts. The mean Brier and Log scores (defined below), again calculated over all questions and forecasts, were $-0.179$ and $-0.560$, respectively. The mean Brier and Log scores (averaging over questions) for the median forecasts (computed within each question) were $-0.138$ and $-0.438$, respectively, whose higher values reflect the ``crowd wisdom" phenomenon that follows from the use of concave scoring functions.

\subsection{Scoring aggregated forecasts}

To evaluate forecasts and aggregates of forecasts for binary event outcomes we consider positively oriented Brier and Log scores,
\begin{equation}
    \label{eq:BrierScore}
    \begin{aligned}
        S_{\text{Brier}}(X,p) &= -(p - X)^2\\[0.05in]
        S_{\text{Log}}(X,p) &= X\log(p) + (1-X)\log(1-p)
    \end{aligned} 
\end{equation}
where $p \in [0,1]$ denotes a forecast for the outcome variable $X$ which takes value zero if the forecast question resolves as ``no" and one if it resolves as ``yes". We contextualize these raw scores, with subscripts removed so as to refer to either Brier or Log, using a skill score,
\begin{equation}
S_{\text{Skill}}(X,p,p_0) = \frac{S(X,p) - S(X,p_0)}{S(X,X) - S(X,p_0) },
\end{equation}
where $p_0$ is a benchmark or reference forecast and the $S(X,X)$ in the denominator is the optimal score, given by a perfect ``oracle" forecaster, which is zero for the Log and Brier scores. The skill score serves to shift and scale the raw scores so that a skill score of zero represents no improvement over the benchmark and a skill score of one represents unimprovable forecasting ability. A comprehensive and authoritative discussion of probability scores can be found in \cite{gneiting2007strictly} where it is noted that, in general, skill scores are not \textit{strictly proper} unless $S(X,p_0)$ is independent of outcome, which holds in the binary case only if $p_0=0.5$. Thus they should not generally be used as rewards in forecasting competitions. Nevertheless, in the context of retrospective analyses they remain a useful tool for making comparisons between forecasters and forecasting methodologies.

Our final stage of score processing involves the aggregation of skill scores over the time interval during which forecasts can be made. Specifically, for each question and each  aggregation method we compute skill scores for the aggregation of forecasts made prior to certain times. We have chosen to use three of these, equidistant in calendar time, that we associate with early-, middle- and late-stage forecasts. Weighted and unweighted mean averages of the skill scores at these three times are reported in Table \ref{tab:brier_scores} (described in more detail below), where the weights are linearly decreasing in calendar time so that early, prescient forecasts are rewarded more highly.

\subsection{Performance Evaluation}

To assess the effectiveness of kairosis, we compare its performance against three competitor methods. Each of the these methods (1)-(4) can be considered as providing a weighting for aggregating individual forecasts. A universal forecast of $0.5$ provides a more primitive comparison.
\begin{enumerate}
\item a uniform weighting ({\em i.e.\ }leading to unweighted aggregate forecasts);
\item a kairosis-weighting (with five  bins, $\alpha_{k} = \lambda \CP, \alpha'_{k} = 1$, $\lambda = 0.2$ concentration parameters, and $p = 1/10$ in the geometric decay prior);
\item a binary weighting that effectively discards the oldest 80\% of forecasts (as proposed by \citealt{himmelstein2023wisdom});
\item a weighting that decays exponentially in forecaster time (the decay rate being specified to match the prior distribution for the change point location used in the kairosis-weighting);
\item a universal forecast of $0.5$ for all questions.
\end{enumerate}
As noted earlier, kairosis can be seen as being positioned between methods (3) and (4). A likelihood function that is constant over time, which is approached if the $\{\alpha_k,\alpha_k'\}$ were to be made very large, makes for a posterior distribution for the change point location that is the same as the exponential prior. The exponential prior then leads to exponentially decaying forecast aggregation weights. Alternatively, by sending the parameter $p$ to zero the exponential prior becomes flat and by sending $\{\alpha_k,\alpha_k'\}$ to zero we heighten our sensitivity to the bin counts so that large changes (which may or may not fall around the 80\% mark) result in a step-like CMF and a step-like weighting function. 

In each case (except (5), where they are equal) we compute both the weighted mean and the weighted median (that is, for forecasts $i=1,\ldots,R$ ordered by increasing weights $w_i$, the forecast $f_j$ with $j$ the smallest integer such that $\sum_{i=1}^j w_i > {1 \over 2}$).

In Table \ref{tab:brier_scores} we present the four different aggregated skill scores (using Log and Brier raw scores, and uniform and time-decreasing weights) for the eight different forecast aggregation methods averaged across the 650 questions, along with the universal forecast of $0.5$. The skill scores use the unweighted median forecast aggregation as the benchmark, so that the entries in the top row of the table are all necessarily zero. (Note that the universal $0.5$ performs vastly worse than all the aggregation methods.) The best (highest) score in each column is placed in a box, and on each of the four measures this is the kairosis-weighted median. Notice that the unweighted mean performs worse than benchmark on all scores. The kairosis median is the only method to perform better than benchmark on all four scores.

\newpage

\begin{landscape}
\begin{table}[ht!]
\centering
\begin{tabular}{@{}lllllll@{}}
                                   &                    & &  \multicolumn{4}{l}{Aggregate skill scores }                                          \\ \cmidrule(l){4-7} 
                                   &                    & 
                                   \multicolumn{1}{l}{\textbf{Brier scores}} &
                                   \multicolumn{2}{l}{From raw Brier scores} &
                                   \multicolumn{2}{l}{From raw Log scores} \\ 
                                   \cmidrule(l){4-7} 
 &
   &
  \begin{tabular}[c]{@{}l@{}}Unweighted\\ Brier scores\end{tabular} &
  \begin{tabular}[c]{@{}l@{}}Unweighted\\ over time\end{tabular} &
  \begin{tabular}[c]{@{}l@{}}Weighted\\ over time\end{tabular} &
  \begin{tabular}[c]{@{}l@{}}Unweighted\\ over time\end{tabular} &
  \begin{tabular}[c]{@{}l@{}}Weighted\\ over time\end{tabular} \\
Forecast weighting                 & Forecast aggregate &                      &                    &                    &                    &                    \\ \cmidrule(r){1-2}
\multirow{2}{*}{Uniform}           & Median & -0.143 (0.006) & 0.000 (0) & 0.000 (0) & 0.000 (0) & 0.000 (0) \\ \cmidrule(lr){2-2}
                                   & Mean & -0.147 (0.005) & -0.657 (0.136) & -0.652 (0.135) & -0.199 (0.022) & -0.198 (0.022) \\ \cmidrule(r){1-2}
\multirow{2}{*}{Kairosis}          & Median & \fbox{-0.137 (0.006)} & \fbox{0.060 (0.009)} & \fbox{0.054 (0.009)} & \fbox{0.046 (0.006)} & \fbox{0.042 (0.006)} \\ \cmidrule(lr){2-2}
                                   & Mean & -0.140 (0.005) & -0.540 (0.134) & -0.545 (0.133) & -0.139 (0.022) & -0.143 (0.022) \\ \cmidrule(r){1-2}
\multirow{2}{*}{Most recent 20\%}  & Median & -0.143 (0.006) & -0.135 (0.067) & -0.159 (0.069) & -0.011 (0.015) & -0.023 (0.015) \\ \cmidrule(lr){2-2}
                                   & Mean & -0.144 (0.005) & -0.694 (0.347) & -0.726 (0.357) & -0.146 (0.029) & -0.157 (0.030) \\
                                   \cmidrule(r){1-2}
\multirow{2}{*}{Exponential decay} & Median & -0.151 (0.006) & -0.211 (0.034) & -0.252 (0.038) & -0.040 (0.015) & -0.061 (0.017) \\ \cmidrule(lr){2-2}
                                   & Mean & -0.149 (0.006) & -0.447 (0.075) & -0.476 (0.077) & -0.124 (0.018) & -0.138 (0.019) \\
\cmidrule(r){1-2}
\multirow{1}{*}{All Forecasts 0.5}  & All Forecasts 0.5 & -0.250 (0.000) & -18.658 (5.681) & -18.525 (5.677) & -2.170 (0.198) & -2.154 (0.197) \\
\end{tabular}
\caption{\textbf{Performance comparison for forecast aggregation methods.} Scores for aggregated forecasts are averaged over 650 forecast questions and three forecast times (early-, middle- and late-stage). Rows index methods for forecast aggregation and columns index variants of the skill score. Table entries are skill scores benchmarked against the unweighted median, so that positive values indicate better-than-benchmark performance and negative values worse-than-benchmark performance. For each skill score variant (i.e. each column) the best forecast is boxed. Standard deviations are presented in parentheses.}
\label{tab:brier_scores}
\end{table}

\end{landscape}
\newpage

\begin{landscape} 
\begin{figure}[t]
    \begin{tabular}{cc}
        \begin{subfigure}{0.668\textwidth}
            \includegraphics[width=\linewidth]{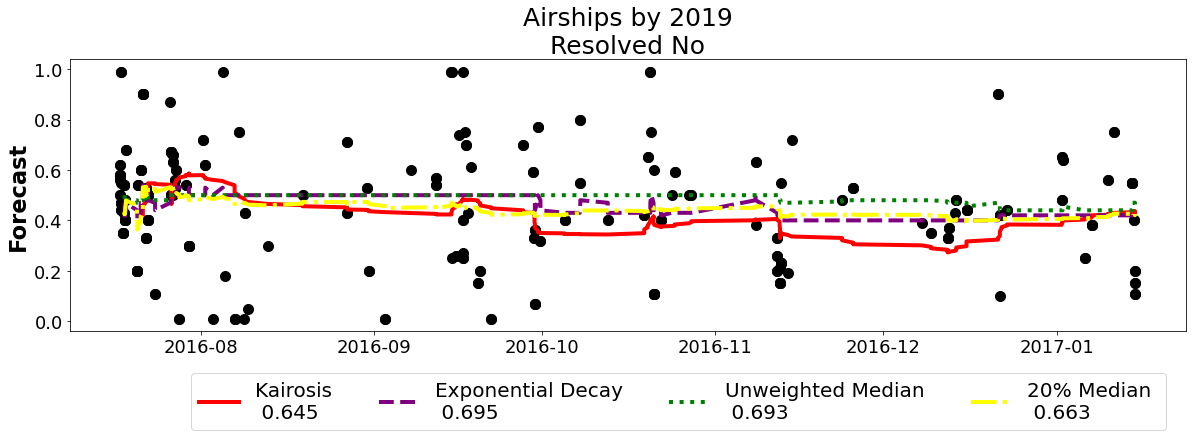}
            \caption{}
        \end{subfigure} &
        \begin{subfigure}{0.668\textwidth}
            \includegraphics[width=\linewidth]{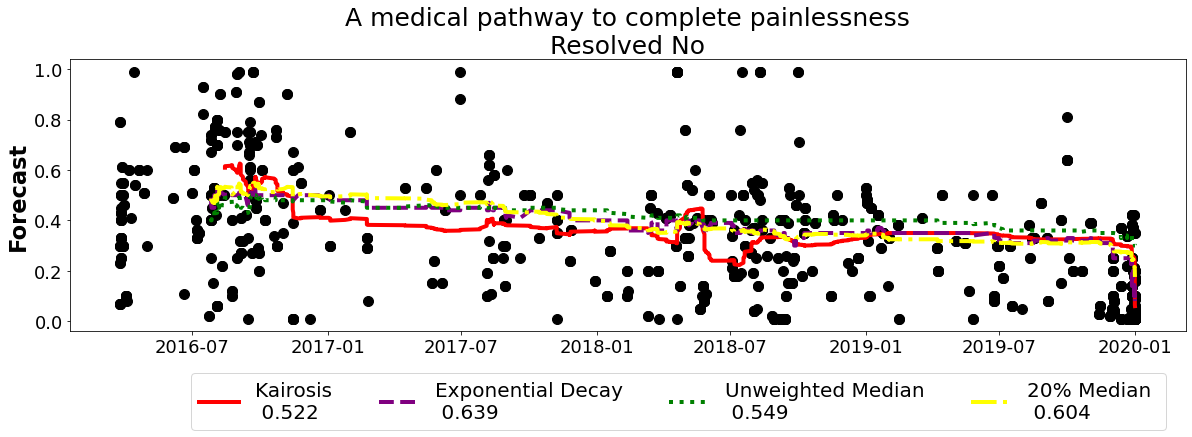}
            \caption{}
        \end{subfigure} \\
        \begin{subfigure}{0.668\textwidth}
            \includegraphics[width=\linewidth]{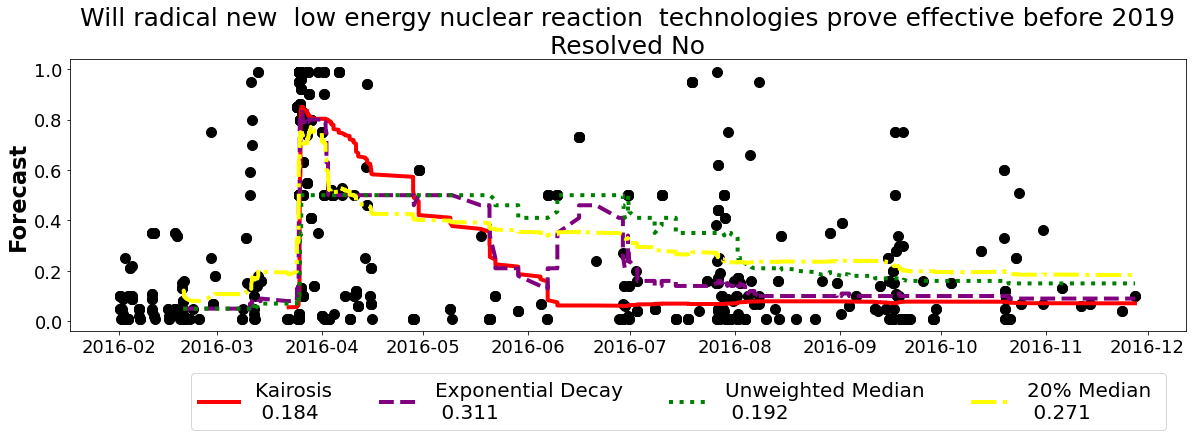}
        \caption{}
        \end{subfigure} &
        \begin{subfigure}{0.668\textwidth}
            \includegraphics[width=\linewidth]{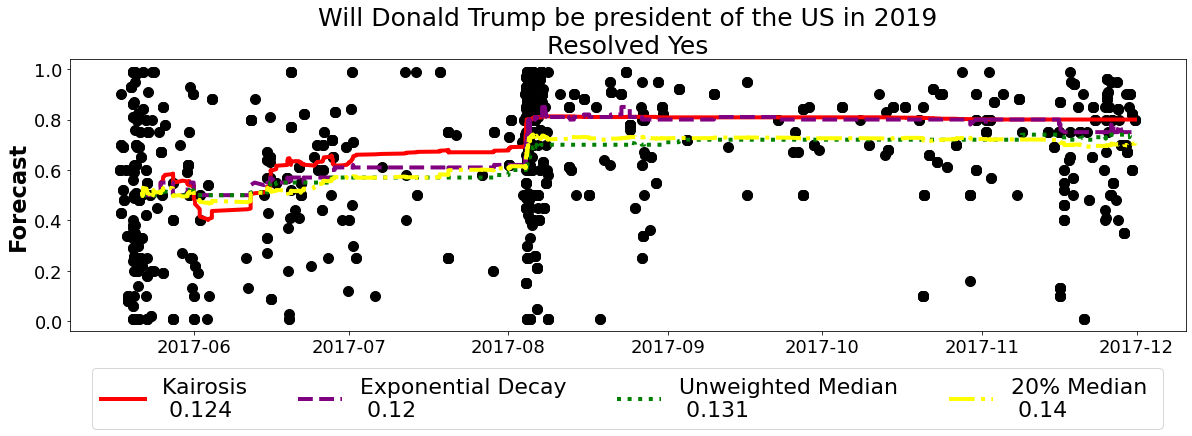}
            \caption{}
        \end{subfigure} \\
       
        \begin{subfigure}{0.668\textwidth}
            \includegraphics[width=\linewidth]{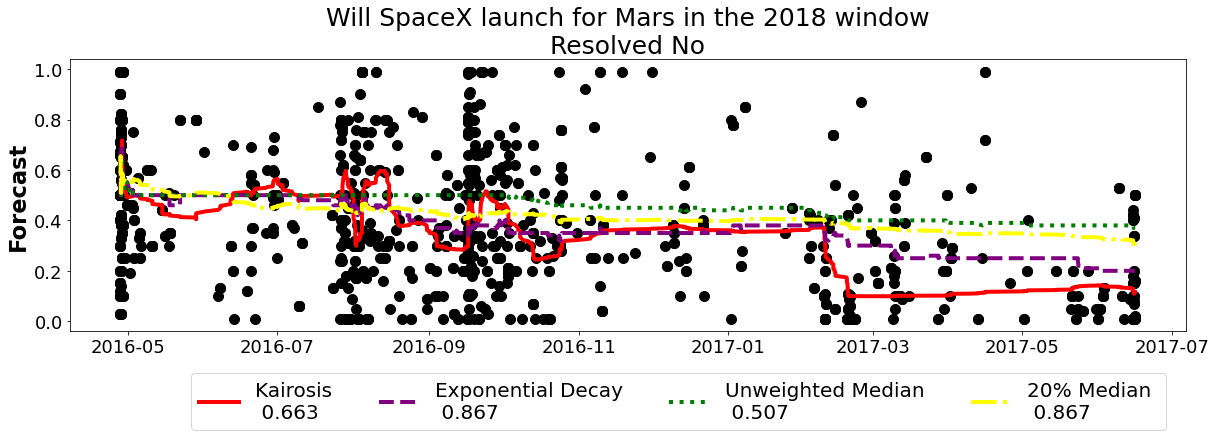}
            \caption{}
        \end{subfigure} &
        \begin{subfigure}{0.668\textwidth}
            \includegraphics[width=\linewidth]{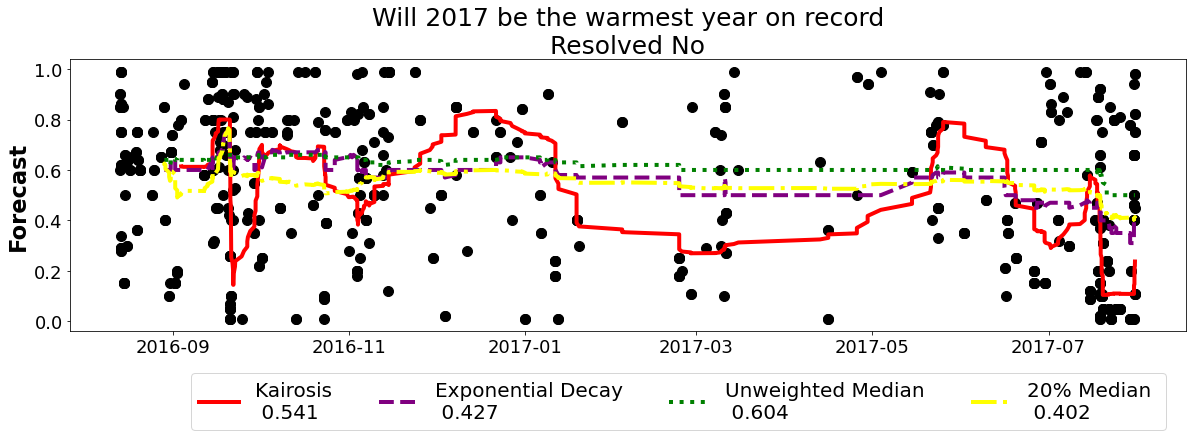}
            \caption{}
        \end{subfigure} \\
        \begin{subfigure}{0.668\textwidth}
            \includegraphics[width=\linewidth]{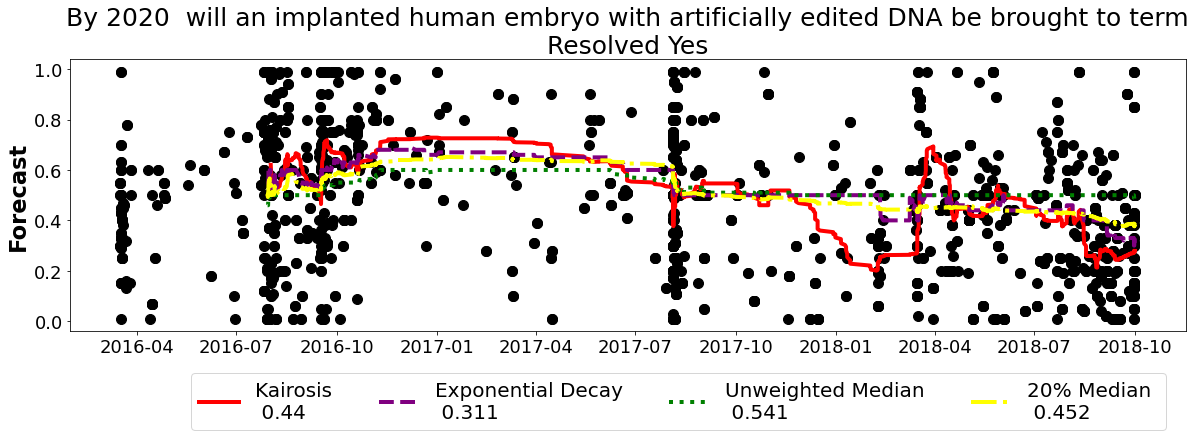}
            \caption{}
        \end{subfigure} &
        \begin{subfigure}{0.668\textwidth}
            \includegraphics[width=\linewidth]{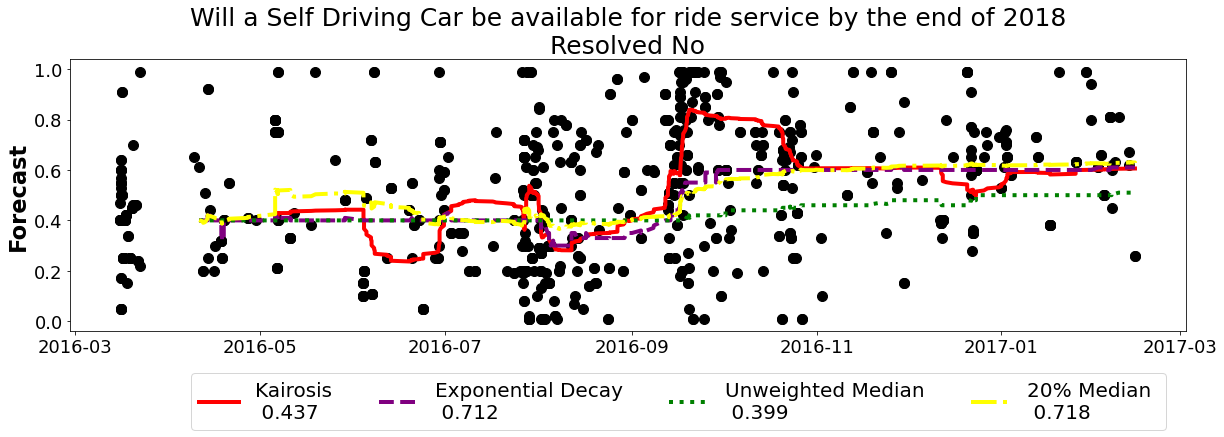}
            \caption{}
        \end{subfigure} \\

    \end{tabular}
    \caption{\textbf{Aggregation methods for a subset of questions are examined}. Aggregated (median) forecasts using different weighting schemes are computed for a subset of questions from the Metaculus dataset. These forecasts, computed at each time point, are interpolated and plotted. Log scores, also computed at each time point and weighted according to time before the question resolution, are averaged and recorded in the figure legends.}
    \label{fig:forecast comparisons}
\end{figure}
\end{landscape}

\subsection{Remarks on kairosis and crowd inaccuracy}

The existence of shared biases within a crowd of forecasters imposes a natural limit on the effectiveness of any aggregate forecast. In the context of the Metaculus questions, the shared bias can be attributed to the forecasters' all inferring event probabilities from subsets of a common, and ultimately inconclusive, set of relevant data. In this sense the forecasts available to us ought to be considered partial observations of the common data rather than of the event itself. Kairosis allows us to adapt to shifting information landscapes but clearly cannot estimate event outcomes to arbitrary accuracy. 

The two key questions now are whether the crowd of Metaculus forecasters possesses a significant amount of information relevant to a particular question, and whether a significant proportion of that information can be exploited via kairosis but not with simpler aggregates. The answers vary appreciably between questions. In Figure \ref{fig:forecast comparisons} we take a closer look at the operational dynamics of kairosis on the crowd forecasts as compared with the other methods for eight questions. Sub-figures (a)-(d) illustrate instances where kairosis adapts effectively to directional changes in crowd opinion which prove to be correct. We observe in each case the capacity of kairosis to react quickly but stably to significant changes, and also to respond to steady trends. Conversely, sub-figures (e)-(h) demonstrate scenarios where kairosis accurately follows the crowd's movements, even though the crowd itself was incorrect. Such cases are often characterized by the exaggerated movements of an uncertain crowd (which are tracked by kairosis), perhaps due to overreaction or news events not adding real information, even as a question's closing date approaches.

\section{Discussion}

Our results indicate the broad feasibility of using change-point methods for dynamical probability forecast aggregation. They do so by identifying points in time at which the distribution of forecasts changes, which we attribute primarily to new information emerging and informing the views of the forecasters. The concept of kairos helps us attribute meaning to the change point CMF that provides weights for our aggregated forecasts, linking a classical concept to a computational method by way of a Bayesian model.

Looking forward, we believe it would be particularly useful to combine our work with more sophisticated change point detection methods, specifically online methods such as those proposed in \cite{adams2007bayesian}. A sequential, online approach in which the change point CMF and/or the forecast aggregation weights are updated rather than recomputed may be necessary to scale up and speed up our calculations to larger systems. We suspect that such an approach is also key to generalizing our method to situations in which multiple change points are identified, a problem which otherwise threatens a combinatorial explosion in the number of likelihood evaluations. Many smaller variations to the kairosis methodology also possible, which we invite the reader to experiment with using the Python code provided in the Supplementary Materials. For instance, we used kairos-informed weights to construct a weighted median, but  other measures of central tendency could be used instead. We used only raw forecast data, but measures of forecaster skill could easily be used, say, to weight the counts appearing in \eqref{DCmassfun} so that our change point calculations are more sensitive to the most skilled. 

Kairosis can also be adapted to forecasts on non-binary domains. In \ref{non_prob_appendix} we test the concept by applying kairosis to point forecasts in Metaculus questions on continuous domains: in this form, kairosis works better than other methods, but by a smaller margin. Kairosis could also be used to analyse distributional forecasts of continuous variables where the probabilities are elicited via a set of bins.

The change point model underlying kairosis can also be used as a means to analyze the development of opinions among a population without directly linking this to the outcome of a forecasting question. This would be particularly useful if we were interested, for example, in quantifying the effect of certain events on those opinions. Another possibility would be to infer the degree of collective wisdom of the crowd from its dynamic behaviour, potentially via its under- or over-responsiveness to current events, which could then be used to discount the crowd altogether in favour, say, of some baseline event probability. It would be particularly interesting to reconcile this idea with recent work on herd dynamics among forecasters \citep{keppo2024bayesian}.

In summary, kairosis is an effective new method for dynamic probability forecast aggregation. It is built from a coherent, tractable underlying Bayesian model and makes no assumptions on the distribution of forecasts. We have demonstrated its potential to quickly account for sudden changes in the beliefs of a population of forecasters and anticipate that its good performance will translate to other domains, such as psephology and marketing, involving sentiment and behaviour tracking.

\section*{Acknowledgments}
The authors would like to thank Metaculus for providing data, and Nikos Bosse (Metaculus Research Coordinator) and Christian Williams (Metaculus Director of Communications and Data) in particular for helpful comments and suggestions.

\nocite{*}
\bibliographystyle{plainnat}
\bibliography{references} %
\newpage
\appendix

\section{A worked example of computing the kairosis weighting function}\label{simmed_example_sect}

In this example we step through our kairosis calculations for an artificial example, whose simulated data has been designed to exemplify the types of phenomena we are interested in. For the sake of exposition the example involves only three candidate points in forecaster time for the most recent change point $\CP$. For each candidate value we compute (up to proportionality) a posterior probability according to
\begin{align*}
\underbrace{P(\CP=t_r \mid \text{Forecasts})}_{\text{posterior}}
\propto 
\underbrace{P(\text{Forecasts} \mid \CP=t_r)}_{\text{likelihood}}\underbrace{P(\CP=t_r)}_{\text{prior}}
&&
t_r \in \{t_1,t_2,t_3\}
.
\end{align*}
We specify a prior probability that decays geometrically as explained in Section \ref{deriving_distr_sect}. The likelihood term factorizes into two parts on the basis that a change point is assumed to force a change in the distribution of the forecasts up to and after it so that
\begin{align*}
\hspace*{-0.1in}P(\text{Forecasts} \mid \CP=t_r) = & P(\text{Forecasts made up to }t \mid \CP=t_r) \\
& \times P(\text{Forecasts made after }t \mid \CP=t_r).
\end{align*}
The factors here are based on the bin counts up to and after the proposed change point $t_r$. They are computed using the probability mass function of the Dirichlet-categorical distribution whose functional form is provided in equation \eqref{DCmassfun}. This mass function is maximized when all the counts fall within a single bin and minimized when they are evenly distributed across them all. Intermediate values are attained for intermediate degrees of forecast dispersion.

\begin{figure}[h!]
    \centering
    \begin{subfigure}[t]{0.3\textwidth}
        \centering
        \includegraphics[height=1.2in]{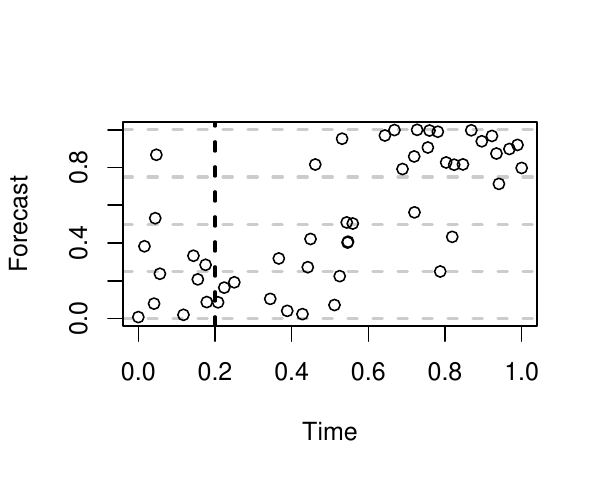}
        \caption{Simulated forecasts partitioned at $t=t_1$.}
    \end{subfigure}\hfill%
    \begin{subfigure}[t]{0.3\textwidth}
        \centering
        \includegraphics[height=1.2in]{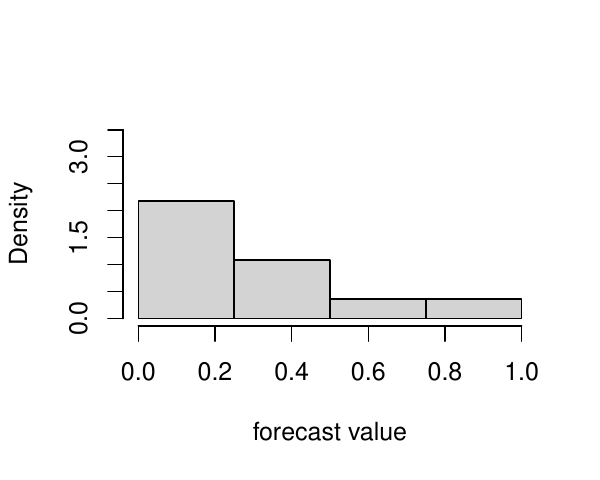}
        \caption{Empirical distribution of pre-change point forecasts for $t=t_1$.}
    \end{subfigure}\hfill
        \begin{subfigure}[t]{0.3\textwidth}
        \centering
        \includegraphics[height=1.2in]{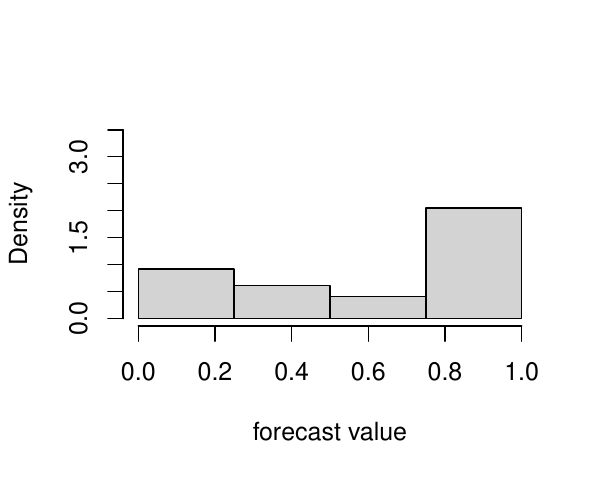}
        \caption{Empirical distribution of post-change point forecasts for $t=t_1$.}
    \end{subfigure}
        \begin{subfigure}[t]{0.3\textwidth}
        \centering
        \includegraphics[height=1.2in]{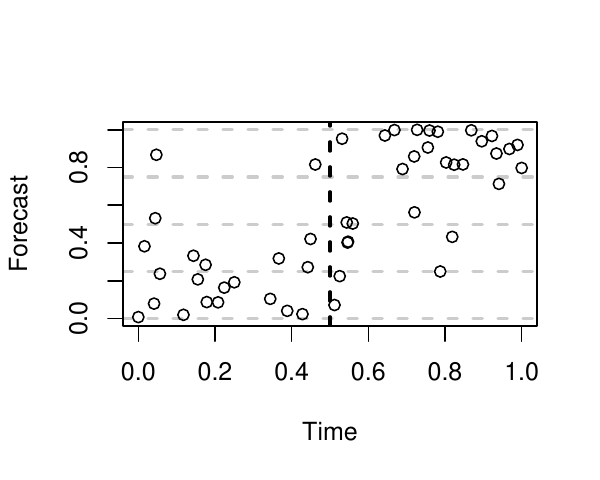}
        \caption{Simulated forecasts partitioned at $t=t_2$.}
    \end{subfigure}\hfill
    \begin{subfigure}[t]{0.33\textwidth}
        \centering
        \includegraphics[height=1.2in]{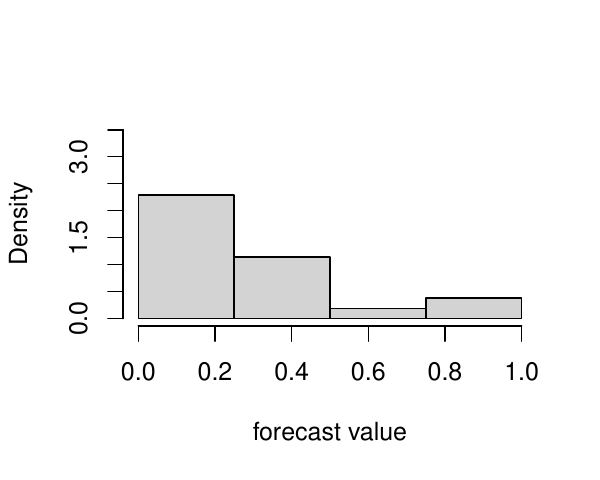}
        \caption{Empirical distribution of pre-change point forecasts for $t=t_2$.}
    \end{subfigure}\hfill
        \begin{subfigure}[t]{0.3\textwidth}
        \centering
        \includegraphics[height=1.2in]{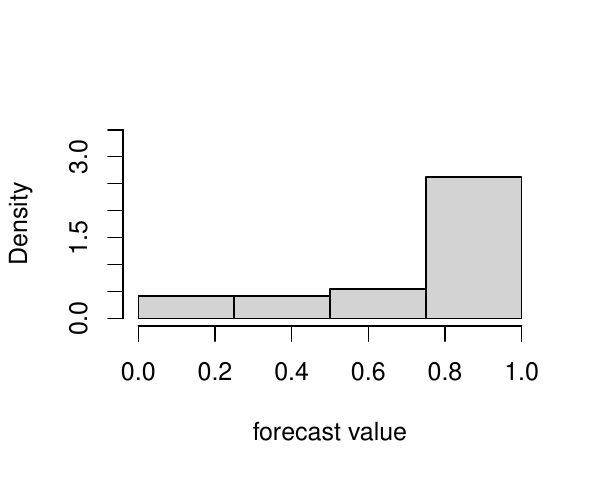}
        \caption{Empirical distribution of post-change point forecasts for $t=t_2$.}
    \end{subfigure}
        \begin{subfigure}[t]{0.3\textwidth}
        \centering
        \includegraphics[height=1.2in]{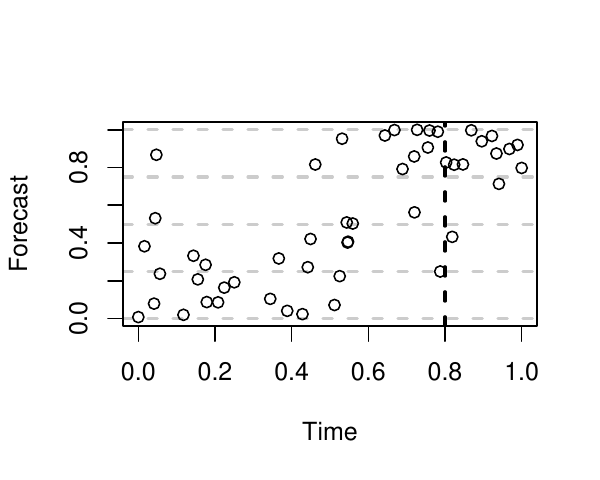}
        \caption{Simulated forecasts partitioned at $t=t_3$.}
    \end{subfigure}\hfill
    \begin{subfigure}[t]{0.3\textwidth}
        \centering
        \includegraphics[height=1.2in]{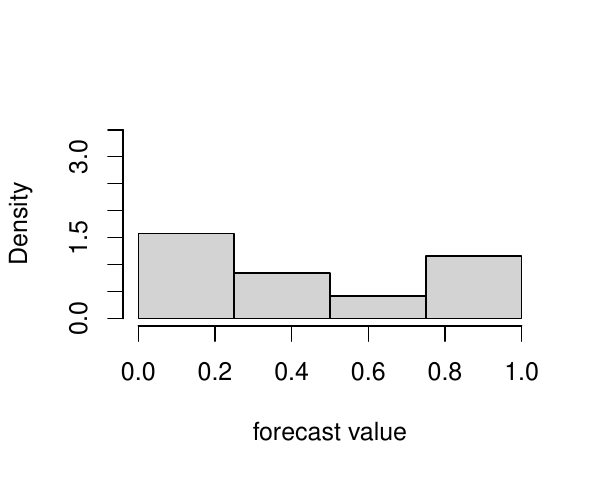}
        \caption{Empirical distribution of pre-change point forecasts for $t=t_3$.}
    \end{subfigure}\hfill
        \begin{subfigure}[t]{0.3\textwidth}
        \centering
        \includegraphics[height=1.2in]{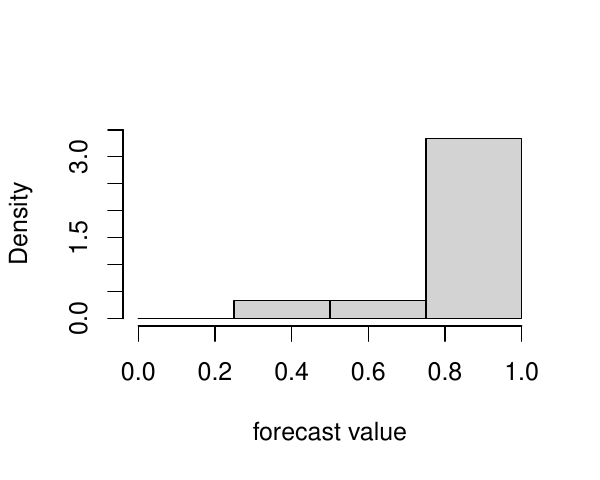}
        \caption{Empirical distribution of post-change point forecasts for $t=t_3$.}
    \end{subfigure}
    \caption{Plots to accompany the illustrative kairosis calculation of \ref{simmed_example_sect}. }\label{simmed_example_plots}
\end{figure}

In Figure \ref{simmed_example_plots} we plot histograms of the pre- and post-change point bin counts for the three candidate time points in our example. The log-likelihoods induced by these counts, which combine additively when computing the log-posterior, are largest when the candidate change point splits the forecasts into subsets both of which are concentrated on a small number of bins. Importantly, these do not need to be the same bins for both the subsets. The log-likelihood is smaller when a set of forecasts is more evenly distributed over the bins.

The values that contribute to the posterior cumulative mass function, which provides us with a weighting function for aggregation, can be found in Table \ref{tab:simmed_example_tab}. We can see, for example, that the earliest and latest candidate change points are assigned very small posterior probabilities. This is owing to the empirical distribution of forecasts on one side of these time points containing a substantial number of both high and low forecasts, leading to relatively large negative log likelihoods (of $-50.667$ and $-51.985$). The greatest posterior probability is calculated for the second candidate time point, which succeeds in partitioning the forecasts into two subsets that are both reasonably concentrated, a feature quantified by the correspondingly moderate contributions (of $-26.321$ and $-34.430$) to the log likelihood.

\begin{figure}[h!]
\centering
\includegraphics[width=0.75\linewidth]{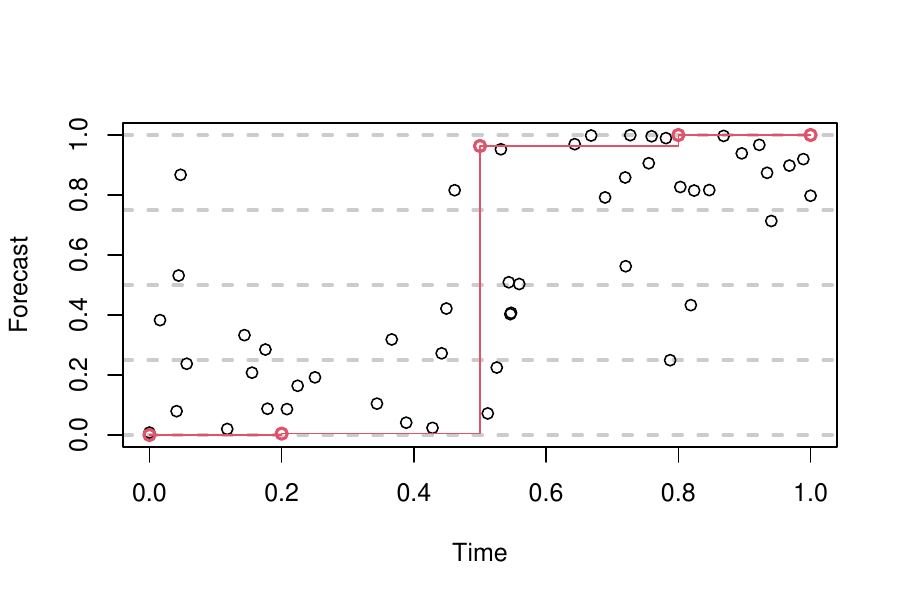}
\caption{The weighting function for forecast aggregation derived as part of our simulated example in \ref{simmed_example_sect}.\label{fig:simmed_example_weight_fun}}
\end{figure}

The three candidate change points lead us to a piecewise linear weighting function with three step discontinuities, as shown in Figure \ref{fig:simmed_example_weight_fun}. When a candidate time point is assigned a large posterior probability of being a change point the values of the weighting function to either side are pushed apart, making the weights lower to the left and higher to the right.

\begin{table}[h!]
\centering
\begin{tabular}{rrrr}
  \hline
 & t= 0.2 & t= 0.5 & t= 0.8 \\ 
  \hline
Log-prior & -5.290 & -4.644 & -3.547 \\ 
  Loglike. for earlier forecasts & -14.792 & -26.321 & -51.985 \\ 
  Loglike. for later forecasts & -50.667 & -34.430 & -13.125 \\ 
  Unnormalized log-posterior & -70.749 & -65.395 & -68.657 \\ 
  Posterior mass funct. & 0.005 & 0.959 & 0.037 \\ 
  Posterior cumulative mass funct. & 0.005 & 0.963 & 1.000 \\ 
   \hline
\end{tabular}
\caption{Intermediate calculations accompanying  \ref{simmed_example_sect}.\label{tab:simmed_example_tab}}
\end{table}

\section{Non-Probabilistic Questions}\label{non_prob_appendix}

As an initial test of the applicability of kairosis beyond probability forecasting for true/false questions, we also applied our method  to  Metaculus questions which required point forecasting of results on continuous domains. This preliminary analysis involved minimal changes to kairosis, but the results are promising. 

The crucial alteration of the method is in the binning of forecasts. Probability forecasts are all on the same, fixed interval $[0,1]$, whereas with general point forecasts the domain varies from question to question, necessitating bins which vary not only between questions but also with present (forecaster) time $R$. In our experimental calculations, whose results are presented below, we adopted a simple adaptive binning strategy which involved partitioning the range of possible forecast values according to the quantiles of the forecasts observed up until the aggregated forecast is needed. This partition defines the bin structure that is used to compare forecast distributions to either side of each candidate change point which, following the calculations described in the main paper, leads to our kairosis weighting function used for aggregation.

Across more than 200 questions with an average open time of 85 days and 2321 forecasts per question, kairosis consistently outperformed both the benchmark and other competing methods. However, it is worth noting that the standard deviations in this scenario are considerably larger than those observed in the probability forecasting questions. Specifically, we observe only a two-standard-deviation difference from the benchmark, in contrast to the more substantial margins seen in probability forecasting. Table \ref{tab:brier_scores} presents a comparison of the unweighted and weighted skill scores, along with their standard deviations, for each of the proposed methods.

\begin{table}[]
\centering
\begin{tabular}{@{}llll@{}}
                                   &                    & \multicolumn{2}{l}{Aggregate skill scores}                                          \\ \cmidrule(l){3-4} 
                                   &                    & \multicolumn{2}{l}{From raw Brier (neg. squared error) scores} \\ \cmidrule(l){3-4} 
 &
   &
  \begin{tabular}[c]{@{}l@{}}Unweighted\\ over time\end{tabular} &
  \begin{tabular}[c]{@{}l@{}}Weighted\\ over time\end{tabular} \\
Forecast weighting                 & Forecast aggregate &                      &                    \\ \cmidrule(r){1-2}
\multirow{2}{*}{Uniform}           & Median & 0.000(0) & 0.000(0) \\ \cmidrule(lr){2-2}
                                   & Mean & -1.459 (0.481) & -1.363 (0.458) \\ \cmidrule(r){1-2}
\multirow{2}{*}{Kairosis}          & Median & \fbox{0.042 (0.027)} & \fbox{0.047 (0.031)} \\ \cmidrule(lr){2-2}
                                   & Mean & -2.655 (0.926) & -2.467 (0.884) \\ \cmidrule(r){1-2}
\multirow{2}{*}{Most recent 20\%}  & Median & -0.709 (0.377) & -0.728 (0.387) \\ \cmidrule(lr){2-2}
                                   & Mean & -3.848 (1.329) & -3.366 (1.079) \\
                                   \cmidrule(r){1-2}
\multirow{2}{*}{Exponential decay} & Median & -0.098 (0.129) & -0.068 (0.091) \\ \cmidrule(lr){2-2}
                                   & Mean & -2.695 (0.933) & -1.363 (0.849) \\
\cmidrule(r){1-2}
\end{tabular}
\caption{Performance comparison for forecast aggregation methods averaged over 200 non-probability forecast questions. Table entries are skill scores benchmarked against the unweighted median, so positive values indicate better-than-benchmark performance, and negative values indicate worse-than-benchmark performance. For each skill score variant, the best forecast is boxed. Standard deviations are presented in parentheses.}
\label{tab:NPbrier_scores}
\end{table}

\section{Sensitivity Analysis}\label{sensitivity_appendix}

\begin{figure}[ht!]
    \centering
        \centering
        \includegraphics[width=\textwidth]{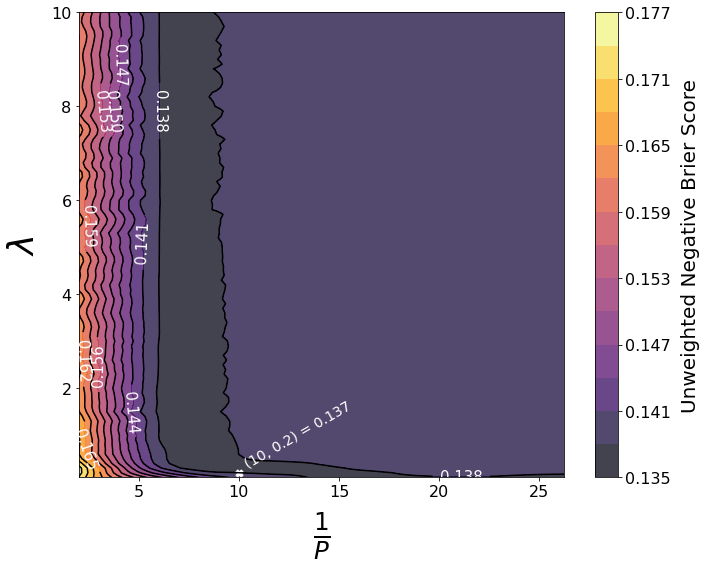}
    \caption{Contour plot showing the sensitivity of the model parameters \(p\) and \(\lambda\) in the kairosis model. The contour plots illustrate how changes in parameter values affect model behavior.}
    \label{fig:sensitivity_analysis}
\end{figure}

We now assess the sensitivity of our kairosis calculations to changes in the parameters $p$, which governs the geometric prior on change point locations, and $\lambda$, which governs the rate at which pseudo-counts $\mathbb{\alpha}_k = \lambda \CP$ are added to the Dirichlet prior for bin probabilities before the supposed most recent change point. Specifically, we calculate Brier scores for aggregated forecasts using weights derived from our kairosis calculations and average these over the Metaculus questions. The calculations are repeated for a range of values of $\{p,\lambda\}$ and the results plotted in Figure \ref{fig:sensitivity_analysis}.

We observe high levels of robustness for all $\lambda > 0.1$ and $p < 0.2$. Larger values of $p$ encode strong prior beliefs that change points occur very often so that the prior probabilities for locations of the last change point decay extremely quickly. The result is an aggregator that places the majority of its weight on only the most recent few forecasts. In this sense the dangers of mis-specifying $p$ are the same as those for mis-specifying the decay parameter for an aggregator using a simple exponential weighting function. We also observe some (less severe) loss in performance for very small values of $\lambda$. Indeed, when $\lambda$ approaches zero we lose some robustness to large movements in the distribution of forecasts before the proposed time of the last change point. We see that the kairosis method is most successful when, via a non-zero specification of $\lambda$, it treats the recent past and the distant past differently.

\end{document}